\documentclass[lettersize,journal]{IEEEtran}
\usepackage{amsmath,amsfonts}
\usepackage{algorithmic}
\usepackage{array}
\usepackage[caption=false,font=footnotesize]{subfig}
\usepackage{textcomp}
\usepackage{stfloats}
\usepackage{url}
\usepackage{verbatim}
\usepackage{graphicx}

\usepackage{algorithm}

\graphicspath{{Fig/}}

\usepackage{cite}
\begin{document}

\title{Joint Time–Phase Synchronization for Distributed Sensing Networks via Feature-Level Hyper-Plane Regression}

\author{Kailun Tian, Kaili Jiang, Dechang Wang, Yuxin Zhao, Yuxin Shang, Hancong Feng and Bin Tang
\thanks{This work was supported in part by the National Natural Science Foundation of China under Grant 62301119. (Corresponding author: Kaili Jiang.)}
\thanks{Kailun Tian, Kaili Jiang, Dechang Wang, Yuxin Zhao, Yuxin Shang, Hancong Feng and Bin Tang are with the University of Electronic Science and Technology of China, Chengdu, Sichuan, 611731, China (e-mail: kailun\_tian@163.com; jiangkelly@uestc.edu.cn; c13844033835@163.com; 1051172535@qq.com; 202411012307@std.uestc.edu.cn; 2927282941@qq.com; bint@uestc.edu.cn).}%
}



\maketitle

\begin{abstract}
Achieving coherent integration in distributed Internet of Things (IoT) sensing networks requires precise synchronization to jointly compensate clock offsets and radio-frequency (RF) phase errors. Conventional two-step protocols suffer from time–phase coupling, where residual timing offsets degrade phase coherence. This paper proposes a generalized hyper-plane regression (GHR) framework for joint calibration by transforming coupled spatiotemporal phase evolution into a unified regression model, enabling effective parameter decoupling. To support resource-constrained IoT edge nodes, a feature-level distributed architecture is developed. By adopting a linear frequency-modulated (LFM) waveform, the model order is reduced, yielding linear computational complexity. In addition, a unidirectional feature transmission mechanism eliminates the communication overhead of bidirectional timestamp exchange, making the approach suitable for resource-constrained IoT networks. Simulation results demonstrate reliable picosecond-level synchronization accuracy under severe noise across kilometer-scale distributed IoT sensing networks.
\end{abstract}


\section{Introduction}
\IEEEPARstart{D}{istributed} coherent sensing networks, such as cooperative uncrewed aerial vehicle (UAV) swarms \cite{jiangDistributedUAVSwarm2024, dingDistributedMachineLearning2024} and wireless edge sensor networks \cite{gulatiReviewPaperWireless2022, nurlanWirelessSensorNetwork2022}, have emerged as a transformative paradigm in the modern Internet of Things (IoT) \cite{wenGenerativeAIGenerative2024}. By coherently integrating spatially distributed nodes into a virtual large-scale aperture, these architectures enable scalable network-level sensing and enhanced coherent processing capability \cite{xuDistributedSignalProcessing2025}. However, realizing the full potential of coherent integration demands extreme synchronization across the entire network \cite{yigitlerOverviewTimeSynchronization2020}. Unlike traditional monolithic arrays tethered by rigid hardware backplanes \cite{liHardwareAccelerationMUSIC2022}, distributed IoT nodes are entirely wireless \cite{merloWirelessPicosecondTime2023}, relying on independent local oscillators and radio-frequency (RF) front-ends. Consequently, synchronization imperfections inherently manifest at two distinct physical levels: the macroscopic clock synchronization error representing the absolute timestamp offset in the nanosecond regime, and the microscopic RF initial phase error representing the fractional wavelength mismatch \cite{marreroArchitecturesSynchronizationTechniques2022}. Achieving precise and joint calibration of these spatiotemporal parameters remains a critical bottleneck in deploying macroscopic IoT sensing clusters.

To address these synchronization challenges, extensive research has been conducted \cite{kClockSynchronizationIndustrial2023}. Packet-based protocols operating at the MAC or network layer, such as the Precision Time Protocol (PTP), Network Time Protocol (NTP), and various broadcast-based reference schemes \cite{shimProvisioningHighPrecisionClock2024, cecchinatoBroadcastSubGHzFramework2024}, are widely adopted in standard wireless networks. Building upon these, numerous closed-loop Two-Way Message Exchange (TWME) and consensus-based algorithms have been proposed to jointly estimate clock skew, offset, and dynamic delays \cite{jingTimeSynchronizationDelay2023, jinFrequencyOffsetInformationAided2024, guCooperativeLocalizationUAV2024, jinNovelConsensusBasedDistributed2024}. While these protocols can achieve microsecond to nanosecond clock alignment, they fundamentally rely on frequent, bidirectional packet handshakes. This paradigm consumes massive communication bandwidth, introduces highly random MAC-layer queuing delays, and completely ignores the RF phase. To mitigate these communication overheads and environmental constraints, approaches like partial timestamp exchange mechanisms \cite{wangClockSynchronizationPartial2023} have been explored. Alternatively, physical-layer algorithms—utilizing techniques like pulse-coupled oscillators and unsupervised deep-learning-assisted synchronization \cite{abakasangaUnsupervisedDeepLearningDistributed2023}—have successfully bypassed MAC-layer uncertainties, pushing temporal accuracy into the sub-nanosecond regime. Nevertheless, existing methodologies almost universally adopt a separated two-step calibration paradigm, in which macroscopic clock delays are first estimated, followed by phase compensation based on the estimated delays \cite{marreroArchitecturesSynchronizationTechniques2022}. This sequential strategy becomes unreliable in the microwave regime, where even picosecond-level residual timing errors are significantly amplified by high-frequency carriers, leading to severe phase distortion and degraded coherent processing performance \cite{nasirTimingCarrierSynchronization2016}. 

In many practical IoT applications, synchronization requirements become the primary bottleneck for system deployment \cite{gunPreciseTwoWay2009}. In such cases, the associated computational cost, power consumption, and implementation complexity of synchronization may even exceed those of the sensing task itself. Meanwhile, the miniaturization trend of distributed IoT sensing nodes imposes stringent constraints on computational resources \cite{javedStateoftheArtFutureResearch2024}, making high-complexity synchronization schemes impractical for edge deployment. Therefore, it is essential to develop a robust and low-complexity joint synchronization framework that enables orthogonal decoupling of macroscopic time delays and microscopic spatial phase errors.

Recently, the Dynamic Manifold theory was introduced, providing a promising geometric perspective for analyzing distributed large-aperture arrays \cite{tianGeometricDirectionFinding2026}. By exploiting the kinematic evolution of non-stationary broadband signals, this theory reveals that under non-negligible macroscopic apertures, the high-order expansion of the dynamic generator intensely couples with large spatial delays. This dynamic coupling forcefully twists the manifold trajectory into a highly complex curve. While existing studies have successfully utilized the intrinsic geometric features (e.g., curvature) of this twisted manifold to resist microscopic RF phase errors, they face two critical gaps when applied to distributed IoT synchronization. First, they exclusively focus on spatial phase ambiguity, completely neglecting the macroscopic clock synchronization offsets. Second, directly extracting parameters from this exact high-dimensional manifold requires multivariate matrix pseudoinverse operations and high-order numerical differentiation. This imposes a prohibitive computational burden, strictly limiting its practical deployment on resource-constrained IoT edge nodes.

To bridge this critical theoretical and engineering gap, this paper proposes a Generalized Hyper-plane Regression (GHR) calibration framework customized for distributed IoT networks. Instead of relying on static spatial snapshots, we systematically establish the exact high-dimensional topological mapping of the dynamic hyper-plane. By formulating this high-order geometric transformation as a multivariate matrix regression model, the highly nonlinear space-time-phase coupling problem is translated into a tractable mathematical framework, allowing for the analytical embedding and orthogonal decoupling of both macroscopic clock errors and microscopic phase errors. 

Building upon this generalized geometric foundation, we develop a targeted feature-level distributed synchronization architecture to guarantee practical IoT deployment. By strategically selecting the Linear Frequency Modulated (LFM) signal as the cooperative calibration source, the dynamic order of the framework is strictly truncated. We mathematically prove that this specific dynamic selection forces the computationally expensive multidimensional hyper-plane to elegantly collapse back into a 2D hyper-line, plummeting the algorithmic computational complexity to a strict linear order. Operating over a passive feature-level interaction mechanism, the distributed nodes unidirectionally transmit highly compressed 1D phase trajectories. This architectural innovation eliminates the communication overhead and queuing uncertainties inherent in bidirectional MAC-layer timestamp exchanges, successfully breaking the error floor of conventional two-step schemes to achieve robust, closed-form picosecond-level synchronization across macroscopic IoT apertures.

The remainder of this paper is organized as follows. Section II establishes the distributed signal model and formulates the fatal time-phase coupling dilemma. Section III systematically develops the generalized dynamic hyper-plane geometry and the orthogonal decoupling mechanism under macroscopic apertures. Section IV details the feature-level distributed calibration architecture, the GHR algorithm, and the theoretical operating boundaries. Section V presents comprehensive experimental results to evaluate the proposed framework. Finally, Section VI concludes the paper.

\section{Distributed Signal Model and Small-Aperture Hyper-line Geometry}
\label{sec:signal_model}

\subsection{Distributed Observation Model and Time-Phase Coupling Dilemma}
\label{subsec:observation_model}

Consider a cooperative far-field source broadcasting a wideband non-stationary signal to a distributed IoT sensing network. Let the analytic representation of the source signal be denoted by $s(t)=\exp (j\Phi (t))$, where $\Phi (t)$ characterizes the total instantaneous phase. The instantaneous frequency of the source is defined as the first-order time derivative of the phase:
\begin{equation}
    \omega (t) = \frac{d\Phi (t)}{dt} = \dot{\Phi }(t)
    \label{eq:inst_freq}
\end{equation}

Assume a distributed coherent IoT sensing network consisting of $M$ spatially separated receiving nodes (e.g., uncrewed aerial vehicles or wireless edge sensors). We designate the first node ($m=1$) as the absolute spatio-temporal reference origin. For the $m$-th distributed node, let $\tau_{m}(\theta)$ denote the spatial propagation delay determined by the Direction of Arrival (DOA) $\theta$ and the distributed node geometry.

In such untethered distributed IoT architectures, the independent local oscillators of the receiving edge nodes cannot achieve perfect synchronization. These inherent synchronization imperfections introduce two distinct levels of error:
\begin{enumerate}
    \item \textbf{Macroscopic Clock Synchronization Error ($\Delta T_{m}$):} Representing the absolute time-stamp offset in the nanosecond regime.
    \item \textbf{Microscopic RF Phase Error ($\Gamma_{m}$):} Representing the initial phase mismatch in the radio frequency (RF) front-end, where $\Gamma_{m} \in [-\pi ,\pi )$.
\end{enumerate}

Consequently, the observation signal received at the $m$-th IoT node is a delayed and phase-rotated version of the source signal, strictly formulated as:
\begin{equation}
    x_{m}(t) = \exp \Big( j\big[ \Phi (t - \tau_{m}(\theta ) - \Delta T_{m}) + \Gamma_{m} \big] \Big) + n_{m}(t)
    \label{eq:obs_signal}
\end{equation}
where $n_{m}(t)$ is the additive white Gaussian noise (AWGN). The distributed observation vector can be compactly written as $\mathbf{x}(t) = [x_{1}(t), x_{2}(t), \ldots , x_{M}(t)]^{T} \in \mathbb{C}^{M}$.

Under this mathematical model, any coherent processing of the signal across the IoT network must address the joint compensation of clock synchronization errors and phase errors. Since $\Phi (t)$ encompasses both the high-frequency carrier and the wideband modulation, any residual error $\delta T$ in estimating the macroscopic clock delay $\Delta T_{m}$ will translate into a massive, time-varying phase rotation $\omega(t)\delta T$. This coupling completely randomizes the subsequent estimation of the microscopic phase $\Gamma_{m}$, dictating that the temporal delay ($\tau_{m}+\Delta T_{m}$) and the spatial phase ($\Gamma_{m}$) must be orthogonally decoupled within a joint estimation framework.

\subsection{Dynamic Manifold and Small-Aperture Hyper-line Geometry}
\label{subsec:dynamic_manifold}

To overcome the static ambiguity formulated in Section \ref{subsec:observation_model}, we must abandon the traditional paradigm of treating the time-varying observation $\mathbf{x}(t)$ as a collection of isolated static snapshots. Instead, by exploiting the continuous temporal evolution of the non-stationary signal, we model $\mathbf{x}(t)$ as a smooth, parameterized trajectory tracing a dynamic manifold $\mathcal{M}$ in the high-dimensional complex space $\mathbb{C}^M$ \cite{tianGeometricDirectionFinding2026}.

The local geometric evolution of this manifold is fundamentally characterized by its first-order kinematic derivative, i.e., the tangent vector $\mathbf{v}(t) = \dot{\mathbf{x}}(t)$. Applying the chain rule to the noise-free component of the observation model \eqref{eq:obs_signal}, the element-wise derivative for the $m$-th sensor yields:
\begin{equation}
    v_m(t) = \dot{\Phi}(t - \tau_m(\theta) - \Delta T_m) x_m(t)
\end{equation}
Recalling the definition of instantaneous frequency $\omega(t) = \dot{\Phi}(t)$, the complete tangent vector can be elegantly formulated using a diagonal operator matrix:
\begin{equation}
    \mathbf{v}(t) = \mathbf{\Omega}(\theta, t) \mathbf{x}(t)
    \label{eq:tangent_operator}
\end{equation}
where $\mathbf{\Omega}(\theta, t)$ is the Dynamic Generator, defined as the instantaneous diagonal frequency matrix $\text{diag}(j \omega(t-\tau_{tot,1}), \dots, j \omega(t-\tau_{tot,M}))$, with $\tau_{tot, m} = \tau_m(\theta) + \Delta T_m$ representing the total macroscopic delay.

Equation \eqref{eq:tangent_operator} reveals a profound physical insight: the spatial evolution of the array manifold is strictly driven by the temporal dynamics of the source signal. For traditional microscopic arrays, the delay is negligible, and $\mathbf{\Omega}$ trivially degenerates to a scalar matrix $j \omega(t)\mathbf{I}$. However, in modern distributed sensing networks, the macroscopic spatial separation between nodes, coupled with substantial initial clock synchronization errors, can result in total delays $\tau_{tot, m}$ reaching hundreds of nanoseconds. 

Under such macroscopic scales, the dynamic generator must be expanded using its high-order Taylor series:
\begin{equation}
    \omega(t - \tau_{tot, m}) = \omega(t) - \dot{\omega}(t)\tau_{tot, m} + \frac{1}{2}\ddot{\omega}(t)\tau_{tot, m}^2 - \dots
    \label{eq:generator_expansion}
\end{equation}

Substituting \eqref{eq:generator_expansion} back into the evolution model, it becomes evident that the high-order temporal dynamics of the signal (e.g., $\dot{\omega}(t)$ and $\ddot{\omega}(t)$) explicitly couple with the large delays, acting as severe geometric perturbations. The unknown space-time parameters (DOA $\theta$ and clock offset $\Delta T_m$) are deterministically embedded within the coefficients of these high-order dynamic terms, while the microscopic phase error $\Gamma_m$ is independently isolated as a spatial constant in $\mathbf{x}(t)$.

This kinematic analysis demonstrates that the entangled Space-Time-Phase parameters are systematically encoded within the continuous, high-order evolution of the dynamic manifold. Motivated by this geometric perspective, Chapter 3 will systematically reconstruct this high-dimensional topological mapping and propose the Generalized Hyper-plane Regression framework, aiming to analytically flatten this complex trajectory and achieve exact orthogonal decoupling of all parameters.

\section{Generalized Dynamic Geometry and Hyper-plane Regression Algorithm}
\label{sec:ghr_algorithm}

\subsection{The Dynamic Hyper-plane}
\label{subsec:hyper_plane}

As established in the kinematic analysis of Section \ref{sec:signal_model} and the previous work \cite{tianGeometricDirectionFinding2026}, the spatial phase differences extracted from the zero-order dynamic manifold $\mathbf{x}(t)$ and its first-order tangent vector $\mathbf{v}(t)$ explicitly encapsulate the high-order structural characteristics of the dynamic generator. Note that in Eq. \eqref{eq:tangent_operator}, the tangent vector $\mathbf{v}(t)$ introduces a constant phase shift ($\angle j = \pi/2$) and a frequency-dependent amplitude mapping. Since the instantaneous frequency is strictly positive ($\angle \omega(t) \equiv 0$), this derivative-induced phase shift perfectly vanishes via common-mode cancellation when extracting the relative spatial phase differences between nodes. 

By retaining all high-order terms of the Taylor expansion for the delayed instantaneous phase $\Phi(t - \tau_{tot,m})$, the expansion yields:
\begin{equation}
    \Delta\Psi_m^x(t) = \Delta\Psi_m^v(t) = \sum_{k=1}^{\infty} \frac{(-\tau_{tot,m})^k}{k!} \omega^{(k-1)}(t) + \Gamma_m
    \label{eq:phase_traj_polynomial}
\end{equation}

Equation \eqref{eq:phase_traj_polynomial} reveals a profound topological transformation that directly reflects the kinematic expansion of the dynamic generator. Under macroscopic apertures, the high-order temporal dynamics of the signal---specifically the chirp rate $\dot{\omega}(t)$, the jerk $\ddot{\omega}(t)$, and beyond---intensely couple with the large spatial delays, acting as severe geometric perturbations. By conceptualizing these continuously time-varying dynamic variables $[\omega(t), \dot{\omega}(t), \ddot{\omega}(t), \dots]$ as independent coordinate axes in a high-dimensional parameter space, the twisted complex manifold deterministically flattens onto a generalized \textit{Dynamic Hyper-plane}. 

Mathematically, the generalized dynamic Hyper-plane can be compactly rewritten as a generalized linear observation model:
\begin{equation}
    \Delta\Psi_m^x(t)=\Delta\Psi_m^v(t) = \mathbf{d}^T(t)\mathbf{q}_m + \Gamma_m
    \label{eq:linear_model_compact}
\end{equation}
where the time-varying \textit{Dynamic Basis Vector} $\mathbf{d}(t)$ and the time-invariant \textit{Geometric Parameter Vector} $\mathbf{q}_m$ for the $m$-th node are defined as:
\begin{align}
    \mathbf{d}(t) &= \left[ -\omega(t), \frac{1}{2}\dot{\omega}(t), -\frac{1}{6}\ddot{\omega}(t), \dots \right]^T \label{eq:dynamic_basis} \\
    \mathbf{q}_m &= \left[ \tau_{tot,m}, \tau_{tot,m}^2, \tau_{tot,m}^3, \dots \right]^T \label{eq:geometric_param}
\end{align}

This formulation establishes a universal parameter estimation framework for the dynamic manifold. The continuous dynamics $\mathbf{d}(t)$ explicitly serve as the known non-stationary basis spanning the Hyper-plane subspace, while the macroscopic spatial delays and microscopic phase errors are strictly localized within the constant projection vector $\mathbf{q}_m$ and the intercept $\Gamma_m$, respectively. Consequently, the highly nonlinear Space-Time-Phase coupling problem is formally transformed into a tractable multidimensional linear regression task:
\begin{equation}
    \begin{bmatrix} \hat{\mathbf{Q}} \\ \hat{\boldsymbol{\Gamma}} \end{bmatrix} = \text{Regress}(\boldsymbol{\Psi}, \mathbf{D})
    \label{eq:conceptual_regress}
\end{equation}
The solvability and complexity of this regression strictly depend on the effective dimensionality (i.e., the number of non-zero time-varying components) of the dynamic basis $\mathbf{d}(t)$.

\subsection{Generalized Hyper-plane Regression Algorithm}
\label{subsec:ghr_algorithm}

To accommodate arbitrary non-stationary signals containing high-order dynamic derivatives, we construct a generalized high-dimensional linear regression model, termed the \textit{Generalized Hyper-plane Regression}.

Suppose the processing center gathers continuous valid snapshots over $K$ discrete time indices $\{t_1, t_2, \dots, t_K\}$. For the $m$-th distributed node relative to the reference node, we define the observation vector $\boldsymbol{\psi}_m \in \mathbb{R}^{K \times 1}$ as:
\begin{equation}
    \boldsymbol{\psi}_m = \left[ \Delta\Psi_m^v(t_1), \Delta\Psi_m^v(t_2), \dots, \Delta\Psi_m^v(t_K) \right]^T
    \label{eq:obs_vector_m}
\end{equation}
To process all $M-1$ uncalibrated nodes simultaneously, we construct the global phase observation matrix $\boldsymbol{\Psi} = [\boldsymbol{\psi}_2, \boldsymbol{\psi}_3, \dots, \boldsymbol{\psi}_M] \in \mathbb{R}^{K \times (M-1)}$.

According to the analytical model in \eqref{eq:linear_model_compact}, we stack the time-varying dynamic basis vector $\mathbf{d}(t) \in \mathbb{R}^{d \times 1}$ (where $d$ is the dynamic order of the signal) across the $K$ snapshots to form the known Dynamic Basis Matrix $\mathbf{D} \in \mathbb{R}^{K \times d}$:
\begin{equation}
    \mathbf{D} = \left[ \mathbf{d}(t_1), \mathbf{d}(t_2), \dots, \mathbf{d}(t_K) \right]^T
    \label{eq:basis_matrix}
\end{equation}
Similarly, we construct the global Geometric Parameter Matrix $\mathbf{Q} \in \mathbb{R}^{d \times (M-1)}$ containing the unknown high-order macroscopic delay terms, and the phase error vector $\boldsymbol{\Gamma} \in \mathbb{R}^{1 \times (M-1)}$:
\begin{align}
    \mathbf{Q} &= \left[ \mathbf{q}_2, \mathbf{q}_3, \dots, \mathbf{q}_M \right] \label{eq:Q_matrix} \\
    \boldsymbol{\Gamma} &= \left[ \Gamma_2, \Gamma_3, \dots, \Gamma_M \right] \label{eq:Gamma_vector}
\end{align}

The generalized dynamic Hyper-plane decoupling problem is therefore strictly formulated as a multivariate matrix regression equation:
\begin{equation}
    \boldsymbol{\Psi} = \mathbf{D}\mathbf{Q} + \mathbf{1}_K\boldsymbol{\Gamma} + \mathbf{N}_\Psi
    \label{eq:matrix_regression}
\end{equation}
where $\mathbf{1}_K$ is a $K \times 1$ vector of ones, and $\mathbf{N}_\Psi$ represents the residual Gaussian noise matrix in the phase parameter space.

To decouple the structural geometric matrix $\mathbf{Q}$ from the intercept vector $\boldsymbol{\Gamma}$, we apply the centering projection operator $\mathbf{P}_c = \mathbf{I}_K - \frac{1}{K}\mathbf{1}_K\mathbf{1}_K^T$ to both sides of the equation. This yields the mean-centered matrices $\bar{\boldsymbol{\Psi}} = \mathbf{P}_c\boldsymbol{\Psi}$ and $\bar{\mathbf{D}} = \mathbf{P}_c\mathbf{D}$, effectively eliminating the intercept term:
\begin{equation}
    \bar{\boldsymbol{\Psi}} = \bar{\mathbf{D}}\mathbf{Q} + \mathbf{P}_c\mathbf{N}_\Psi
    \label{eq:centered_equation}
\end{equation}

By utilizing the Ordinary Least Squares (OLS) or Principal Component Analysis (PCA) framework, the optimal estimate for the geometric parameter matrix $\hat{\mathbf{Q}}$ is derived:
\begin{equation}
    \hat{\mathbf{Q}} = (\bar{\mathbf{D}}^T \bar{\mathbf{D}})^{-1} \bar{\mathbf{D}}^T \bar{\boldsymbol{\Psi}}
    \label{eq:Q_estimate}
\end{equation}
Subsequently, the uncalibrated initial phase error vector $\hat{\boldsymbol{\Gamma}}$ is perfectly retrieved by substituting $\hat{\mathbf{Q}}$ back into the centroid of the original observation:
\begin{equation}
    \hat{\boldsymbol{\Gamma}} = \frac{1}{K}\mathbf{1}_K^T \left( \boldsymbol{\Psi} - \mathbf{D}\hat{\mathbf{Q}} \right)
    \label{eq:Gamma_estimate}
\end{equation}

Finally, recognizing that the first row of $\hat{\mathbf{Q}}$ solely contains the first-order total macroscopic delay terms $\tau_{tot,m} = \tau_m(\theta) + \Delta T_m$, the clock synchronization error $\Delta T_m$ can be explicitly decoupled given the spatial geometry $\tau_m(\theta)$. The complete execution flow is summarized in Algorithm \ref{alg:ghr_algorithm}.

\begin{algorithm}[t]
\caption{The GHR Algorithm}
\label{alg:ghr_algorithm}
\begin{algorithmic}[1]
\renewcommand{\algorithmicrequire}{\textbf{Input:}}
\renewcommand{\algorithmicensure}{\textbf{Output:}}
\REQUIRE $\{\boldsymbol{\psi}_m\}_{m=2}^M$, $\mathbf{D}$ and $\boldsymbol{\tau}(\theta)$.
\ENSURE $\{\Delta \hat{T}_m\}_{m=2}^M$ and $\{\hat{\Gamma}_m\}_{m=2}^M$.

\STATE $\boldsymbol{\Psi} = [\boldsymbol{\psi}_2, \boldsymbol{\psi}_3, \dots, \boldsymbol{\psi}_M]$.
\STATE $\mathbf{P}_c = \mathbf{I}_K - \frac{1}{K}\mathbf{1}_K \mathbf{1}_K^T$.
\STATE Centralize matrices: $\bar{\boldsymbol{\Psi}} \leftarrow \mathbf{P}_c \boldsymbol{\Psi}$ and $\bar{\mathbf{D}} \leftarrow \mathbf{P}_c \mathbf{D}$.

\STATE $\hat{\mathbf{Q}} \leftarrow (\bar{\mathbf{D}}^T \bar{\mathbf{D}})^{-1} \bar{\mathbf{D}}^T \bar{\boldsymbol{\Psi}}$.
\STATE $\hat{\boldsymbol{\Gamma}} \leftarrow \frac{1}{K} \mathbf{1}_K^T ( \boldsymbol{\Psi} - \mathbf{D} \hat{\mathbf{Q}} )$.

\STATE $\hat{\boldsymbol{\tau}}_{tot} \leftarrow \hat{\mathbf{Q}}(1, :)$.
\FOR{$m = 2$ to $M$}
    \STATE Decouple clock error: $\Delta \hat{T}_m \leftarrow \hat{\tau}_{tot, m} - \tau_m(\theta)$.
    \STATE Map RF phase error to principal branch $[-\pi, \pi)$: \\
    $\hat{\Gamma}_m \leftarrow \text{wrapToPi}(\hat{\Gamma}_m)$.
\ENDFOR

\RETURN $\{\Delta \hat{T}_m\}_{m=2}^M$ and $\{\hat{\Gamma}_m\}_{m=2}^M$.
\end{algorithmic}
\end{algorithm}

\section{Distributed GHR Calibration Framework and Performance Boundaries}
\label{sec:framework}

Building upon the rigorous mathematical foundation of the GHR algorithm derived in Section \ref{sec:ghr_algorithm}, this section translates the pure multidimensional regression theory into a highly efficient, practical distributed engineering protocol tailored for IoT applications. We propose a passive feature-level interaction architecture, explicitly introduce a dimensionality reduction mechanism to resolve algorithmic complexity bottlenecks on resource-constrained edge devices, and rigorously analyze the fundamental physical boundaries governing the system's performance.

\subsection{Distributed Feature-Level Interaction Architecture}
\label{subsec:interaction_architecture}

Traditional distributed synchronization protocols, such as Two-Way Message Exchange (TWME), fundamentally rely on the bidirectional transmission of discrete MAC-layer packets to extract timestamps. This paradigm consumes massive communication bandwidth, introduces highly random queuing delays, and severely compromises the energy efficiency and electromagnetic stealth of non-cooperative sensing clusters like UAV swarms and wireless edge sensors.

\begin{figure*}[t]
    \centering
    \includegraphics[width=\textwidth]{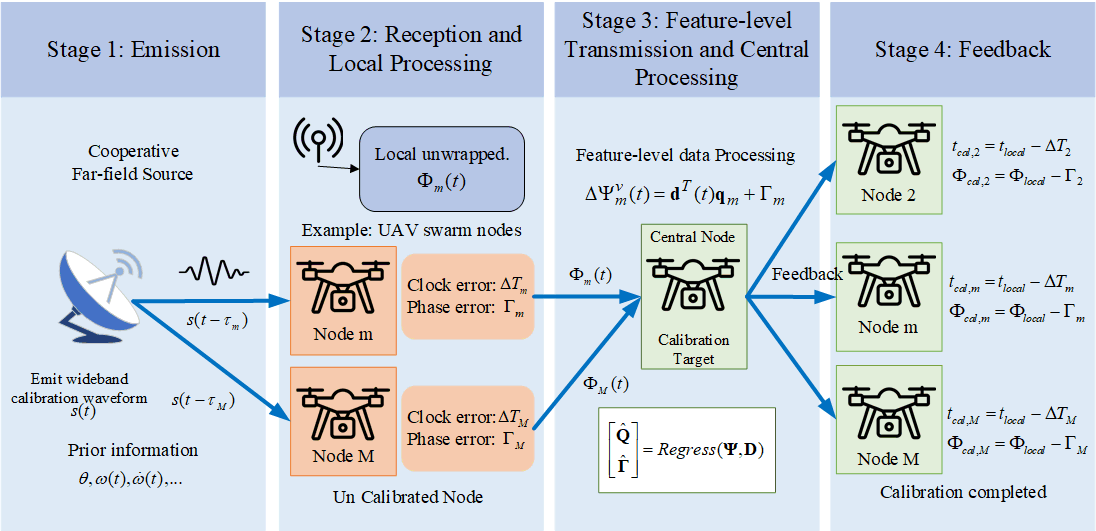}
    \caption{Architecture of the proposed feature-level cooperative calibration framework in a distributed IoT sensing network.}
    \label{fig:Fig1}
\end{figure*}

To overcome these bottlenecks, Fig. \ref{fig:Fig1} illustrates the overall implementation architecture of the proposed distributed GHR calibration framework, which adheres to a strict feature-level interaction paradigm designed for low-bandwidth IoT environments.

In the first stage, a known wideband calibration waveform is broadcast by a far-field cooperative source with a prior angle of arrival $\theta$.

Each distributed IoT node captures this signal and performs robust local processing to extract the unwrapped spatial phase trajectory. It is worth noting that, although the phase of $\mathbf{x}$ can be directly extracted using a single antenna at each node, such estimation is highly susceptible to noise in practical scenarios. To enable robust local phase extraction, we advocate equipping each node with a subarray and performing local processing based on subarray-level tangent vectors $\mathbf{v}$, which significantly enhances system robustness. 

Subsequently, in Stage 3, instead of transmitting gigabytes of raw RF baseband data, each uncalibrated edge node forwards only its one-dimensional, highly compressed feature-level data vector, $\boldsymbol{\psi}_m$, to a central data processing node via a unidirectional link. 

Finally, the central node constructs the gathered phase matrix $\boldsymbol{\Psi}$ and utilizes the GHR algorithm to jointly and orthogonally decouple the clock and RF phase errors, feeding them back to precisely calibrate the distributed swarm.

This architecture thoroughly eliminates the need for bidirectional packet exchanges, bypassing MAC-layer queuing uncertainties while ensuring a kilobyte-level communication footprint. 

\subsection{Algorithmic Complexity and Dimensionality Reduction}
\label{subsec:complexity_lfm}

While the GHR framework provides a universal mathematical model capable of handling arbitrary dynamic orders ($d \ge 1$), performing multidimensional matrix operations poses a severe computational bottleneck for resource-constrained IoT edge networks. Specifically, the computational complexity of the centralized GHR algorithm is exclusively dominated by the pseudoinverse and matrix multiplications. For $M-1$ uncalibrated nodes, $K$ valid temporal snapshots, and a source signal with dynamic order $d$, computing the covariance matrix $\bar{\mathbf{D}}^T \bar{\mathbf{D}}$ requires $\mathcal{O}(K d^2)$ floating-point operations. The subsequent matrix inversion requires $\mathcal{O}(d^3)$ operations, and the cross-correlation projection $\bar{\mathbf{D}}^T \bar{\boldsymbol{\Psi}}$ requires $\mathcal{O}(K d M)$ operations.

Thus, the overall computational complexity scales as $\mathcal{O}(K d^2 + d^3 + K d M)$. For highly non-stationary signals with large dynamic orders, extracting errors from high-dimensional hyper-planes inherently imposes a heavy computational burden, threatening the real-time responsiveness and battery life of IoT nodes.

To guarantee real-time engineering feasibility, we strategically select the LFM signal as the cooperative calibration source. For an LFM signal, the chirp rate is a known constant ($\dot{\omega}(t) = 2\pi\mu$), and all higher-order derivatives identically vanish ($\ddot{\omega}(t) = 0$). This specific dynamic condition enforces a strict truncation of the dynamic order to $d=1$. Most crucially, because $\dot{\omega}(t)$ is constant, the second-order perturbation term ceases to be a time-varying coordinate axis, collapsing into a deterministic scalar. The multidimensional Hyper-plane precisely degenerates into a 2D Hyper-line:
\begin{equation}
\begin{aligned}
\Delta \Psi_m^v(t) &= -\left( \tau_m(\theta) + \Delta T_m \right)\omega(t) \\
&\quad + \underbrace{\left[ \Gamma_m + \pi\mu\left( \tau_m(\theta) + \Delta T_m \right)^2 \right]}_{\text{Generalized Intercept } \tilde{\Gamma}_m}
\end{aligned}
\label{eq:lfm_hyperline}
\end{equation}

Algorithmically, this deliberate selection allows the dynamic basis matrix $\mathbf{D}$ to collapse into a single-column instantaneous frequency vector. The computationally expensive $d \times d$ matrix inversion gracefully degenerates into a trivial scalar division requiring merely $\mathcal{O}(1)$ operations. Correspondingly, the overall computational complexity of the decoupling algorithm plummets to a strict linear order of $\mathcal{O}(KM)$. By driving the computational overhead to its theoretical minimum, the LFM-based framework guarantees robust execution even for massive-scale distributed IoT sensing clusters.

\subsection{Theoretical Operating Boundaries}
\label{subsec:theoretical_boundaries}

The validity of the proposed framework is governed by two fundamental physical boundaries: the algorithmic phase Nyquist limit and the (signal-to-noise ratio) SNR threshold.

The phase unwrapping operation inherently requires that the phase variation between adjacent discrete samples does not exceed $\pm \pi$. For a signal with bandwidth $B$ and pulse duration $T_{dur}$, the maximum frequency variation per sample (at sampling rate $f_s$) is $\Delta f = B / (T_{dur} f_s)$. The corresponding maximum phase increment induced by a macroscopic delay $\tau_{max}$ is $\Delta \phi = 2\pi \Delta f \cdot \tau_{max}$. To avoid catastrophic cycle slips along the temporal axis, we must satisfy $\Delta \phi \le \pi$. This establishes the rigorous maximum unambiguous macroscopic aperture bound:
\begin{equation}
    D_{max} = c \cdot \tau_{max} \le \frac{c \cdot f_s \cdot T_{dur}}{2B}
    \label{eq:nyquist_limit}
\end{equation}
This theoretical boundary mathematically guarantees that higher sampling rates or extended pulse durations linearly expand the resolvable macroscopic distance, ensuring the algorithm's compatibility with kilometer-scale distributed IoT networks.

In addition, the fundamental constraint of the proposed decoupling algorithm lies in the non-linear phase unwrapping process, where severe noise can induce catastrophic phase jumps (i.e., cycle slips), destroying the topological structure of the extracted hyper-plane. Although the cooperative calibration signal typically guarantees a relatively high SNR, investigating the framework's performance under low-SNR conditions remains practically meaningful for resource-constrained edge nodes. 

To overcome this threshold, introducing a local sub-array at each distributed node is highly advocated. For example, the local processing module can perform coherent integration over the sub-array observations, which significantly enhances the anti-noise capability during local phase extraction. Consequently, as long as the post-integration SNR stays above the phase wrapping threshold, the generalized hyper-plane decoupling remains strictly valid. To validate this mechanism, the experimental section of this paper explicitly compares the performance disparity between directly extracting the phase of $\mathbf{x}$ (using a single antenna) and employing the local sub-array processing ($\mathbf{v}$), decisively demonstrating the framework's capability to reliably operate in severely degraded environments.

The local processing module leverages techniques like Singular Value Decomposition to perform coherent integration over a sliding window of length $L$. This optimal rank-1 projection yields an asymptotic processing gain of $10\log_{10}(L)$ dB. The generalized hyper-plane decoupling remains valid as long as the post-integration SNR stays above the non-linear phase unwrapping threshold. Consequently, the minimum required SNR follows a phase transition behavior. While an enlarged distributed aperture necessitates higher geometric stability, the substantial integration gain allows the proposed framework to reliably operate in severely degraded environments (e.g., 0 dB), decisively breaking the performance floor of conventional packet-based synchronization protocols.

\section{Experimental Results}
\label{sec:experiments}
\subsection{Experiment 1: High-Order Manifold Topology and Multidimensional Hyper-plane Validation}
\label{subsec:exp1}

This experiment aims to verify the correctness of the geometry of dynamic manifold hyperplanes on a large, distributed scale. For the distributed receiver nodes, the sampling rate is set to $5\text{ GHz}$. The system consists of two distributed nodes separated by a macroscopic distance of approximately $30$ meters. The uncalibrated node is injected with a clock synchronization error of $3.45\text{ ns}$ and an RF initial phase error of $1.234\text{ rad}$. For the far-field transmitting node, the carrier frequency is $2\text{ GHz}$, the signal pulse width is $1\ \mu\text{s}$, and the bandwidth is unified to $500\text{ MHz}$. Four typical signal types are evaluated: a LFM signal, a Sinusoidal Frequency Modulated (SFM) signal with a modulation rate of $2\text{ MHz}$, a Quadratic Frequency Modulated (QFM) signal, and a 2-level Frequency-Shift Keying (2FSK) signal with a symbol rate of $10\text{ MBaud}$. To isolate the pure geometric morphology from receiver thermal noise, a SNR of $50\text{ dB}$ is applied.

\begin{figure}[t]
    \centering
    \subfloat[LFM signal.]{\includegraphics[width=0.48\linewidth]{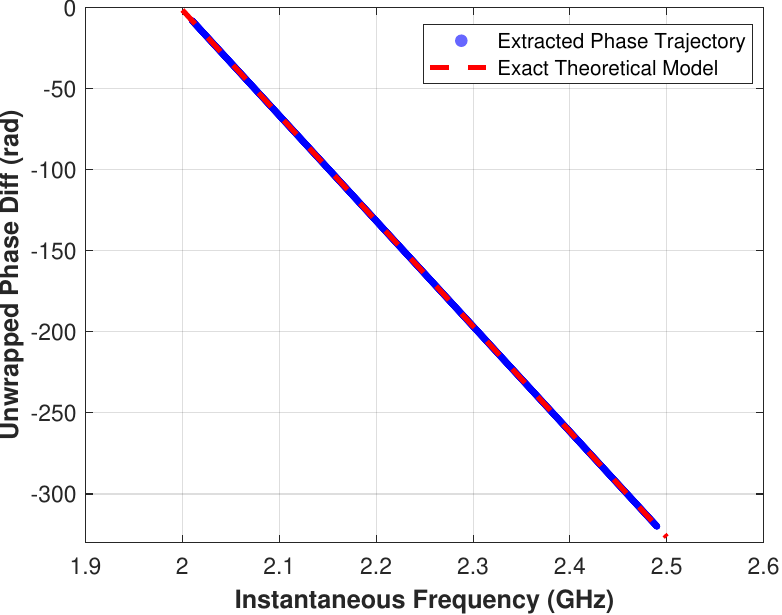}\label{fig:fig2a}}
    \hfill
    \subfloat[SFM signal.]{\includegraphics[width=0.48\linewidth]{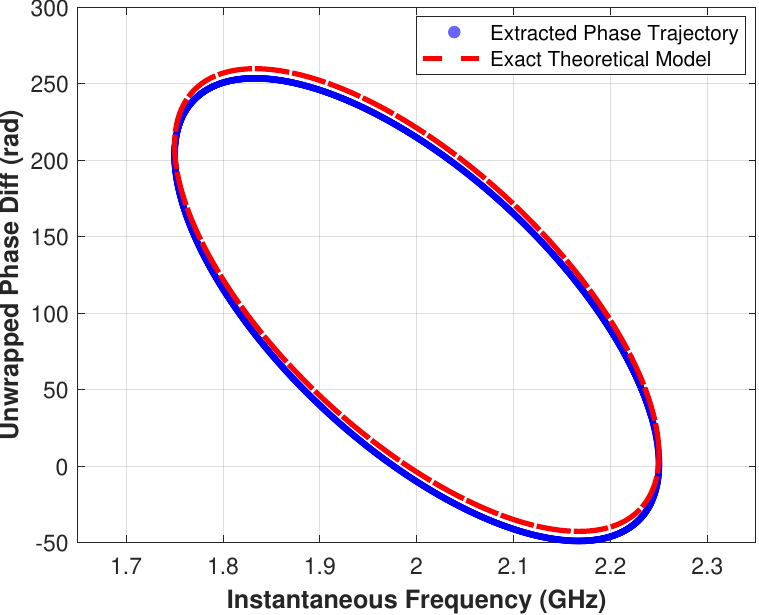}\label{fig:fig2b}}
    \vskip\baselineskip
    \subfloat[QFM signal.]{\includegraphics[width=0.48\linewidth]{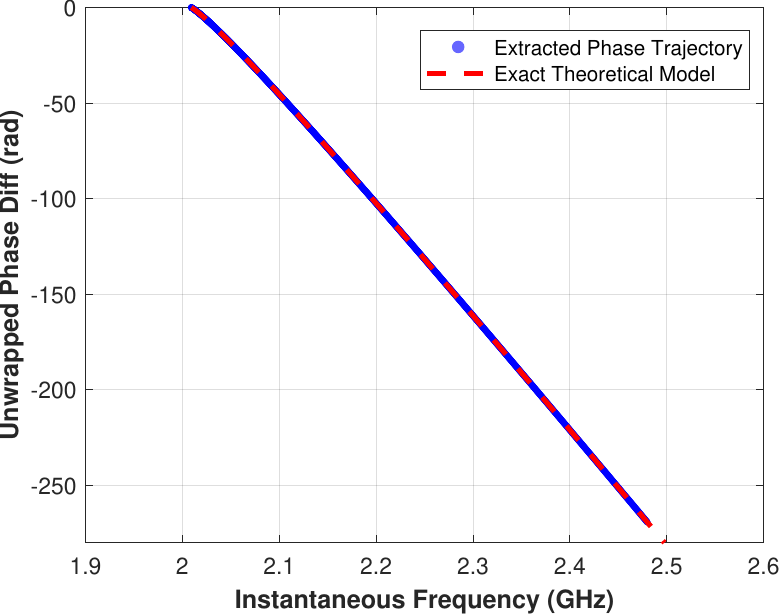}\label{fig:fig2c}}
    \hfill
    \subfloat[2FSK signal.]{\includegraphics[width=0.48\linewidth]{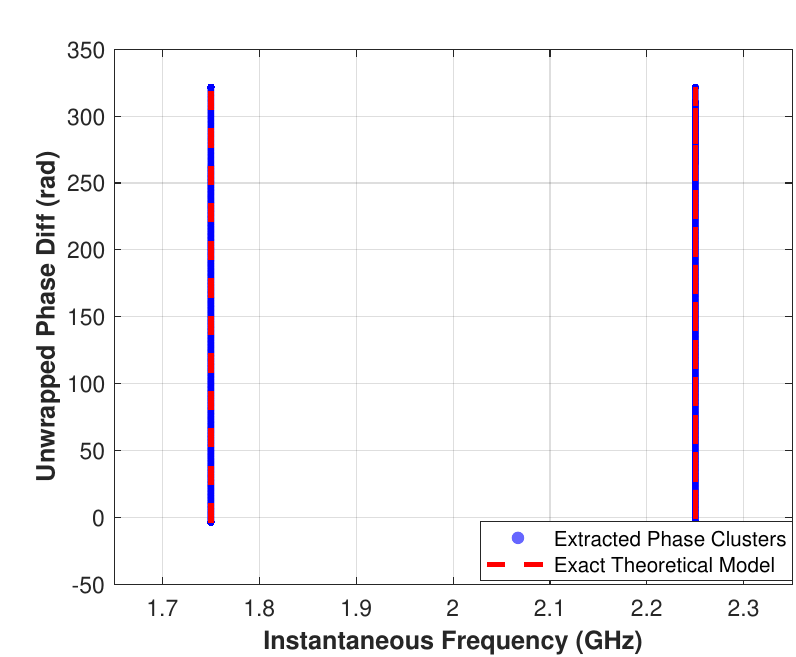}\label{fig:fig2d}}
    \caption{High-order manifold topology of far-field transmitting node signal.}
    \label{fig:fig2}
\end{figure}

\begin{figure}[t]
    \centering
    \includegraphics[width=0.8\linewidth]{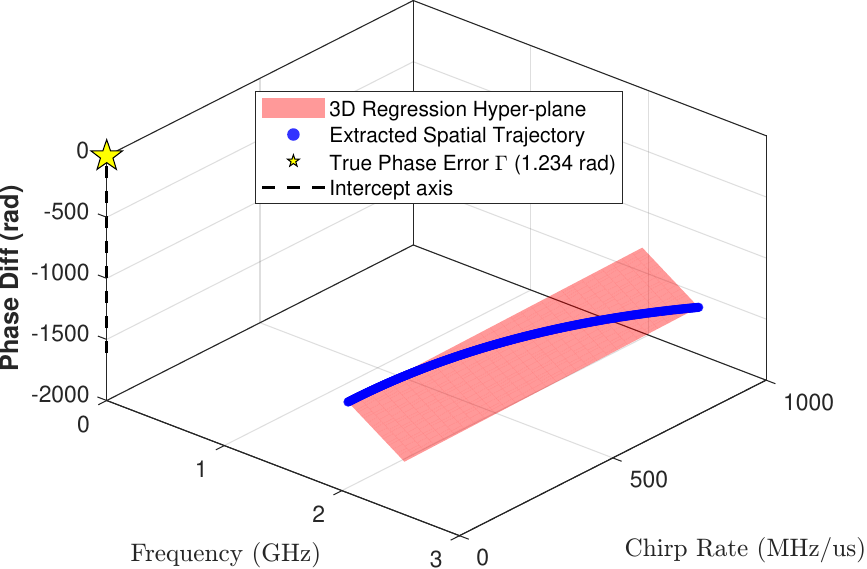}
    \caption{Multidimensional Hyper-plane of QFM signal.}
    \label{fig:fig3}
\end{figure}

Fig.~\ref{fig:fig2} illustrates the geometric topologies of the four signals projected onto the two-dimensional instantaneous frequency versus unwrapped phase difference plane. Under a traditional microscopic array aperture, the second-order delay residuals approach zero, and the phase evolution of all signals presents as a straight line. However, under the macroscopic distributed aperture, the high-order dynamics of the signals explicitly reshape the topological structure of the two-dimensional projection. Since the instantaneous chirp rate of the LFM signal is constant, its second-order residual degenerates into a fixed intercept shift, maintaining a one-dimensional line morphology, as shown in Fig. \ref{fig:fig2}(a). In contrast, as shown in Fig. \ref{fig:fig2}(b), the oscillating chirp rate of the SFM signal distorts its two-dimensional projection into a closed Lissajous ellipse, the linearly varying chirp rate of the QFM signal transforms its projection into a parabola, and the discrete frequency hopping of the 2FSK signal manifests as separated data clusters, as shown in Fig. \ref{fig:fig2}(c) and \ref{fig:fig2}(d). The exact high-order theoretical analytical curves match the extracted observation point clouds, confirming that the macroscopic aperture amplifies the non-linear coupling of high-order dynamics, which invalidates the two-dimensional linear model for signals with non-constant chirp rates.

Fig. \ref{fig:fig3} demonstrates the three-dimensional hyper-plane geometric characteristics of the QFM signal. By expanding the observation phase space to a three-dimensional parameter space including instantaneous frequency and instantaneous chirp rate, the curve that projected in the two-dimensional plane is flattened onto a tilted dynamic hyper-plane. Extrapolating this three-dimensional hyper-plane towards zero frequency and zero chirp rate yields an intercept on the vertical axis that matches the pre-set true RF initial phase error. The estimated clock synchronization error is $3.4508\text{ ns}$, with a calibration accuracy of $7.9863 \times 10^{-4}\text{ ns}$. The estimated phase error is $-1.8966\text{ rad}$, with a calibration error of $3.1306\text{ rad}$. This demonstrates that constructing a multivariate dynamic linear regression model by introducing the chirp rate as an independent variable can absorb the high-order evolution non-linearity of complex signals, achieving topological flattening and orthogonal decoupling of the macroscopic clock and microscopic RF initial phase.

\subsection{Experiment 2: The GHR Algorithm Under LFM Signal}
\label{subsec:exp2}

This experiment aims to verify that, when an LFM signal is employed as the calibration signal, the GHR algorithm can be reformulated as a regression problem over a multidimensional hyper-linear geometric structure, and to further demonstrate that equipping each node with a subarray can provide significant performance gains compared to a single-antenna configuration.

For a distributed system, the sampling rate of each node is set to $5\text{ GHz}$. The system consists of three distributed nodes, where the first node serves as the reference, and the subsequent nodes are separated by macroscopic distances corresponding to relative propagation delays of $15\text{ ns}$ and $25\text{ ns}$. The clock synchronization errors are set to $3.45\text{ ns}$ and $-2.15\text{ ns}$, and the phase errors are $1.234\text{ rad}$ and $-0.876\text{ rad}$. To provide a comparative ablation analysis, the nodes are alternately configured with a sub-array of 4 antennas spaced at half-wavelength and a degraded single-antenna setup. For the far-field transmitting node, the calibration source emits an LFM signal. The carrier frequency is $2\text{ GHz}$, the signal pulse width is $1\ \mu\text{s}$, and the bandwidth is $500\text{ MHz}$. The SNR is set to $5\text{ dB}$.

\begin{figure}[t]
    \centering
    \includegraphics[width=0.8\linewidth]{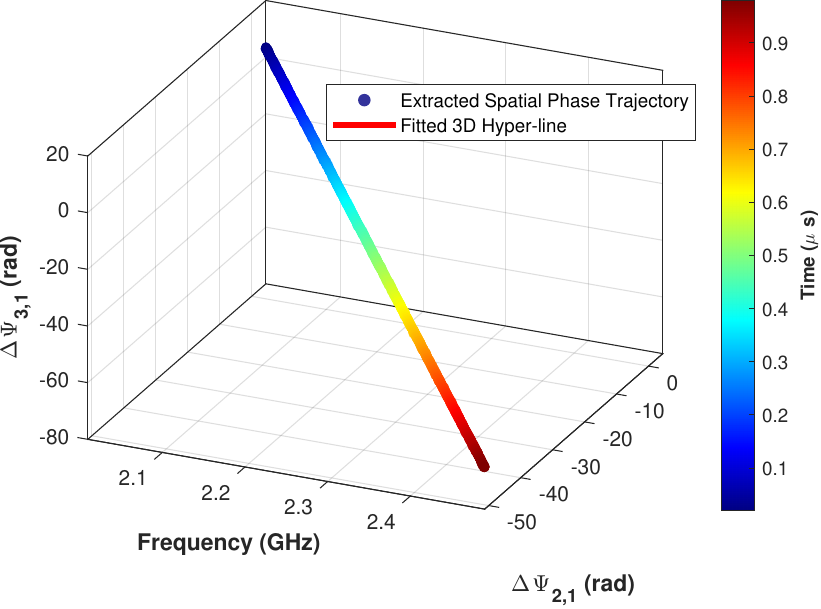}
    \caption{3D Hyper-line Geometry.}
    \label{fig:fig4}
\end{figure}

\begin{figure}[t]
    \centering
    \subfloat[2D projection using four antennas.]{\includegraphics[width=0.48\linewidth]{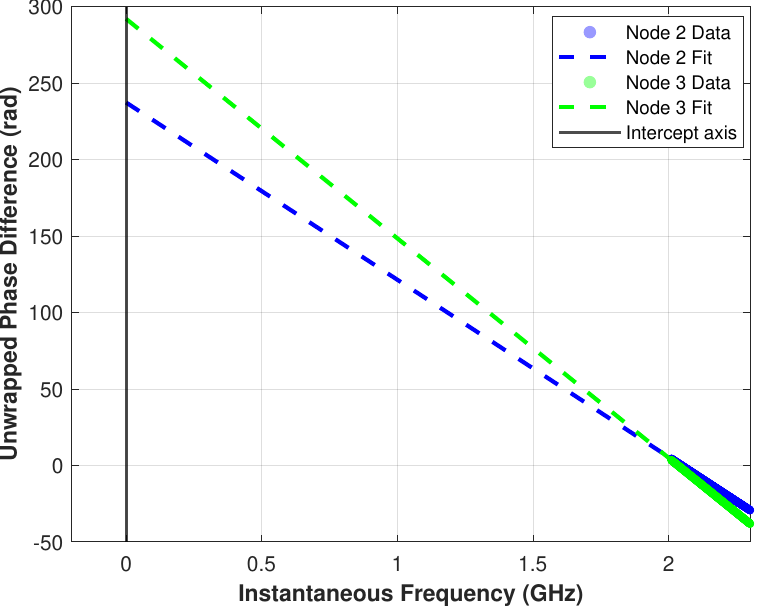}\label{fig:fig5a}}
    \hfill
    \subfloat[2D projection using a single antenna.]{\includegraphics[width=0.48\linewidth]{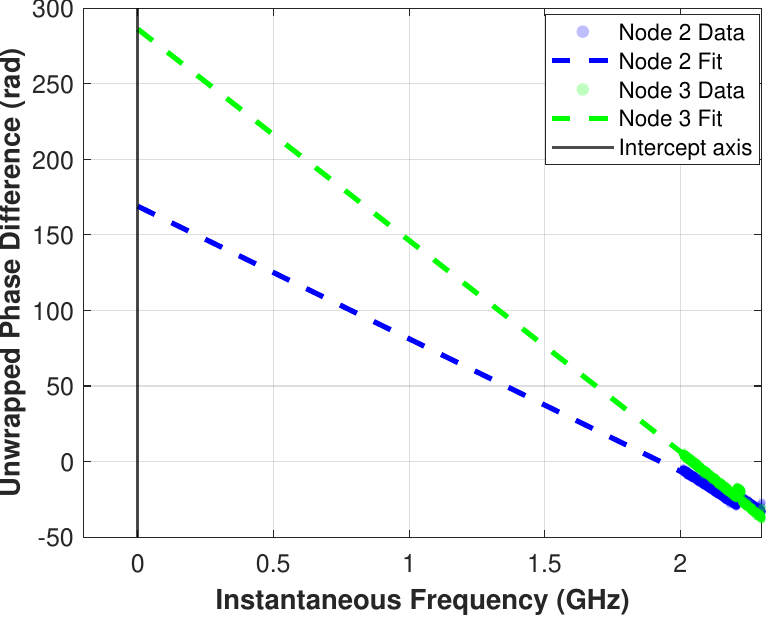}\label{fig:fig5b}}
    \caption{A visual comparison of multi-antenna node's noise-canceling capabilities.}
    \label{fig:fig5}
\end{figure}

Fig. \ref{fig:fig4} illustrates the three-dimensional hyper-line geometry of the three distributed nodes. By extracting the unwrapped spatial phase differences relative to the reference node, the time-evolving dynamic manifold collapses into a structured one-dimensional straight line in the three-dimensional parameter space composed of the instantaneous frequency and dual phase differences. 

Fig. \ref{fig:fig5}(a) presents the two-dimensional projection of the phase differences for the two uncalibrated nodes under the four-antenna configuration at $5\text{ dB}$ SNR. Supported by the local coherent integration of the multi-antenna node, the unwrapped phase trajectories form continuous point clouds that adhere to the theoretical hyper-line model, enabling parameter decoupling. Quantitative analysis of the four-antenna configuration reveals accurate clock error decoupling. For Node 2, the estimated clock error is $3.4474\text{ ns}$ compared to the true value of $3.4500\text{ ns}$ (an error of $2.61 \times 10^{-3}\text{ ns}$), and the estimated phase error is $-1.9458\text{ rad}$ compared to the true $1.2340\text{ rad}$ (an error of $3.18\text{ rad}$). For Node 3, the estimated clock error is $-2.1502\text{ ns}$ compared to the true $-2.1500\text{ ns}$ (an error of $1.71 \times 10^{-4}\text{ ns}$), and the estimated phase error is $2.2622\text{ rad}$ compared to the true $-0.8760\text{ rad}$ (an error of $3.14\text{ rad}$). 

Fig. \ref{fig:fig5}(b) depicts the two-dimensional projection under the single-antenna degradation. Lacking the noise suppression, direct phase unwrapping suffers from cycle slips, causing the point clouds to fracture and diverge significantly from the theoretical linear trend. Consequently, the single-antenna quantitative results deteriorate drastically. For Node 2, the estimated clock error is $-1.0615\text{ ns}$ (an error of $4.51\text{ ns}$) and the estimated phase error is $1.6860\text{ rad}$ (an error of $4.52 \times 10^{-1}\text{ rad}$). For Node 3, the estimated clock error collapses to $-7.8045\text{ ns}$ (an error of $5.65\text{ ns}$) and the estimated phase error is $0.4906\text{ rad}$ (an error of $1.37\text{ rad}$). Evidently, the greater the number of distributed node's internal array elements, the stronger the anti-noise capability.

\subsection{Experiment 3: Validation of Effective Distance for Distributed Nodes}
\label{subsec:exp3}

This experiment aims to validate the effective macroscopic distance between distributed nodes supported by the GHR and analyzes how system parameters influence this operating boundary. For the distributed receiver network, the baseband sampling rate is configured as a variable parameter, tested at $2\text{ GHz}$, $5\text{ GHz}$, and $20\text{ GHz}$. The macroscopic distance between distributed nodes are scanned from $0.1\text{ km}$ up to $15\text{ km}$. For the far-field transmitting node, the calibration source emits an LFM signal. The carrier frequency is $2\text{ GHz}$. The signal pulse duration is tested at $1\ \mu\text{s}$, $10\ \mu\text{s}$, and $20\ \mu\text{s}$. The signal bandwidth is evaluated at $200\text{ MHz}$, $500\text{ MHz}$, and $800\text{ MHz}$. The SNR is fixed at $30\text{ dB}$ to isolate the geometric phase collapse phenomenon from thermal noise interference. The experiment employs a control variable method across three sub-tests to independently evaluate the effects of sampling rate, pulse duration, and signal bandwidth on the maximum resolvable aperture.

\begin{figure}[t]
    \centering
    \includegraphics[width=0.8\linewidth]{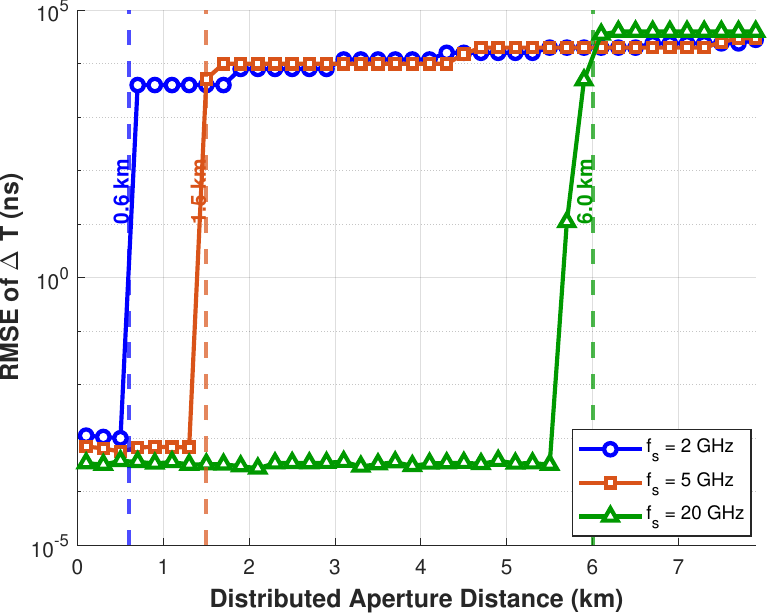}
    \caption{The effect of node sampling rate on distributed effective aperture.}
    \label{fig:fig6}
\end{figure}

As illustrated in Fig. \ref{fig:fig6}, the impact of the hardware sampling rate on the clock error estimation is significant. The root mean square error (RMSE) curves remain stable within the operational range but exhibit a vertical collapse exactly at the theoretical boundaries of $0.6\text{ km}$, $1.5\text{ km}$, and $6.0\text{ km}$, confirming the analytical derivation of the phase Nyquist limit. 

\begin{figure}[t]
    \centering
    \includegraphics[width=0.8\linewidth]{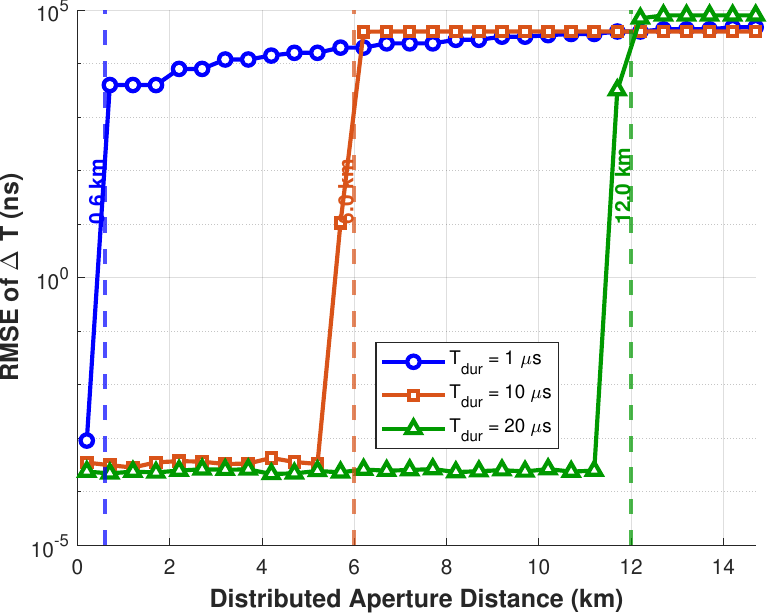}
    \caption{The effect of broadcast node signal duration on distributed aperture.}
    \label{fig:fig7}
\end{figure}

Fig. \ref{fig:fig7} presents the influence of the signal pulse duration. The collapse points match the theoretical limits of $0.6\text{ km}$, $6.0\text{ km}$, and $12.0\text{ km}$, verifying that extended signal durations proportionally expand the maximum unambiguous aperture. 

\begin{figure}[t]
    \centering
    \includegraphics[width=0.8\linewidth]{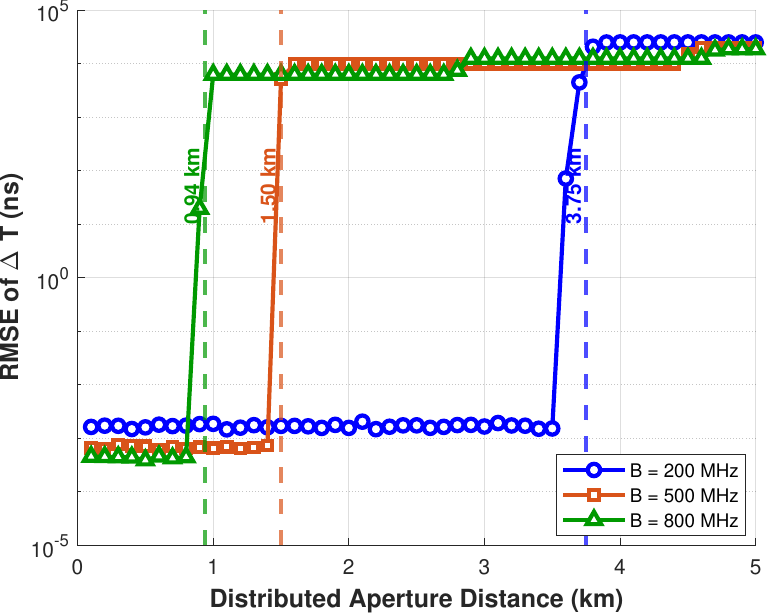}
    \caption{The effect of broadcast node signal bandwidth on distributed aperture.}
    \label{fig:fig8}
\end{figure}

Fig. \ref{fig:fig8} demonstrates the effect of signal bandwidth, where the operational boundaries occur at $3.75\text{ km}$, $1.50\text{ km}$, and $0.94\text{ km}$, showing an inverse relationship with the bandwidth size. 

In addition to the boundary validation, the results within the stable operational zones indicate a clear trend in estimation accuracy. Before reaching the phase collapse boundary, a higher sampling rate at the distributed nodes, a wider signal bandwidth, and a longer pulse duration from the transmitting node jointly contribute to a lower estimation error floor. Conclusively, these findings demonstrate that by appropriately adjusting the system parameters, the proposed scheme can successfully accomplish robust clock synchronization for distributed nodes spanning several kilometers to tens of kilometers, effectively accommodating the scale of modern distributed sensing networks.

\subsection{Experiment 4: Impact of Distributed Aperture Scale on Synchronization Accuracy}
\label{subsec:exp4}

This experiment aims to investigate the impact of the macroscopic distributed aperture size on the estimation accuracy of the clock error and RF initial phase, verifying the immunity of the GHR framework to inter-node distances within the effective phase Nyquist boundary. For the distributed receiver network, a baseband equivalent model is employed with a sampling rate of $5\text{ GHz}$. The network consists of a reference node and an uncalibrated node, alternately configured with a 4-element sub-array and a degraded single-antenna setup. The macroscopic distance offsets between the nodes are evaluated at three distinct levels: $100\text{ ns}$, $500\text{ ns}$, and $1500\text{ ns}$, corresponding to approximate physical separations of $30$ meters, $150$ meters, and $450$ meters. For the far-field transmitting node, the calibration source emits an LFM signal. The carrier frequency is $2\text{ GHz}$, the signal pulse duration is $1\ \mu\text{s}$, and the bandwidth is fixed at $500\text{ MHz}$. The SNR is swept from $-10\text{ dB}$ to $20\text{ dB}$. At each SNR level, 100 independent Monte Carlo trials are performed to obtain statistically stable results.

\begin{figure}[t]
    \centering
    \subfloat[Time calibration RMSE.]{\includegraphics[width=0.48\linewidth]{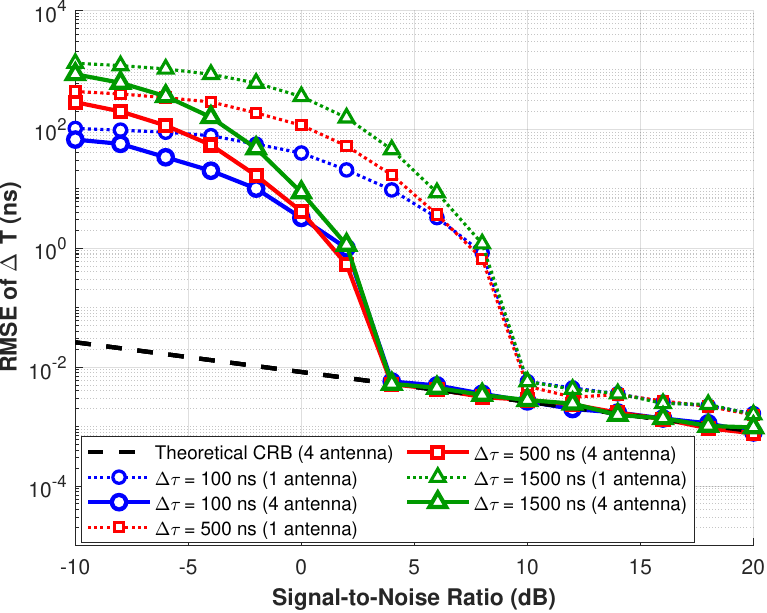}\label{fig:fig9a}}
    \hfill
    \subfloat[Phase calibration RMSE.]{\includegraphics[width=0.48\linewidth]{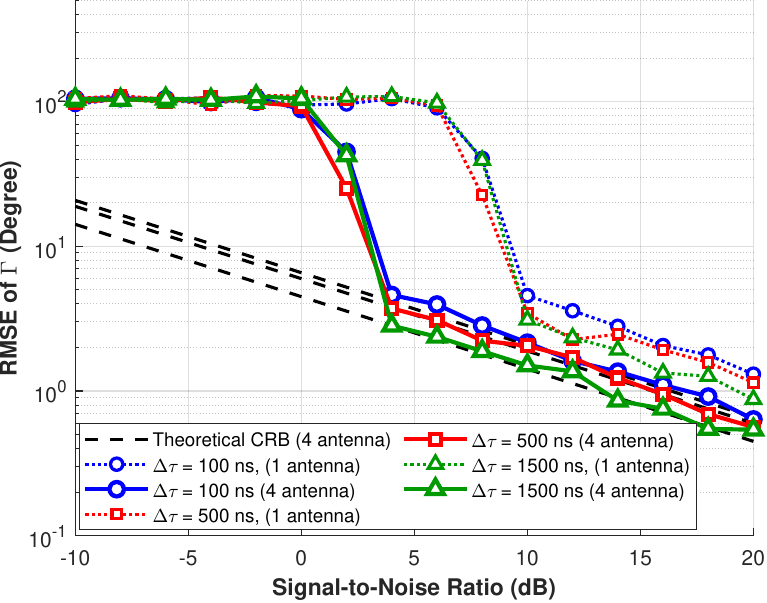}\label{fig:fig9b}}
    \caption{The effect of different spacing patterns within the effective aperture on performance.}
    \label{fig:fig9}
\end{figure}

The simulation results present the RMSE of the clock and phase estimation alongside their corresponding theoretical Cram\'{e}r-Rao Bounds. Fig. \ref{fig:fig9}(a) demonstrates that for the clock error estimation, the theoretical bounds and the proposed GHR curves for all three macroscopic apertures overlap perfectly. This confirms the mathematical derivation that the slope extraction accuracy is completely immune to the macroscopic distance between distributed nodes. Fig. \ref{fig:fig9}(b) reveals the phase error calibration performance. As the aperture scale increases, the second-order phase coupling induces a slight theoretical divergence in the phase Cram\'{e}r-Rao Bounds. Nevertheless, the proposed GHR algorithm strictly tracks these diverging theoretical limits without any performance degradation. In contrast, across both parameter estimations, the single-antenna scheme fails to achieve the theoretical bounds at low and moderate SNRs due to phase wrapping anomalies, emphasizing the necessity of local coherent integration regardless of the aperture scale.

\subsection{Experiment 5: Impact of Signal Bandwidth and Spatial Coherence on Estimation Accuracy}
\label{subsec:exp5}

Building upon the conclusion from Experiment 4 that the macroscopic distance does not fundamentally constrain the synchronization accuracy within operational limits, this experiment fixes a relatively small aperture to investigate the impact of signal bandwidth and spatial coherence on the theoretical estimation limits. For the distributed receiver network, the baseband sampling rate remains at $5\text{ GHz}$. The macroscopic distance between the reference node and the uncalibrated node is fixed at approximately $3$ meters. To conduct an ablation study on spatial coherence, the distributed nodes are alternately configured with a 4-element sub-array and a degraded single-antenna setup. For the far-field transmitting node, the calibration source emits an LFM signal. The carrier frequency is $2\text{ GHz}$, and the signal pulse duration is $1\ \mu\text{s}$. The signal bandwidth is evaluated at three distinct levels: $200\text{ MHz}$, $500\text{ MHz}$, and $800\text{ MHz}$. The SNR is swept from $-10\text{ dB}$ to $20\text{ dB}$, with 100 independent Monte Carlo trials performed at each level.

\begin{figure}[t]
    \centering
    \subfloat[Time calibration RMSE.]{\includegraphics[width=0.48\linewidth]{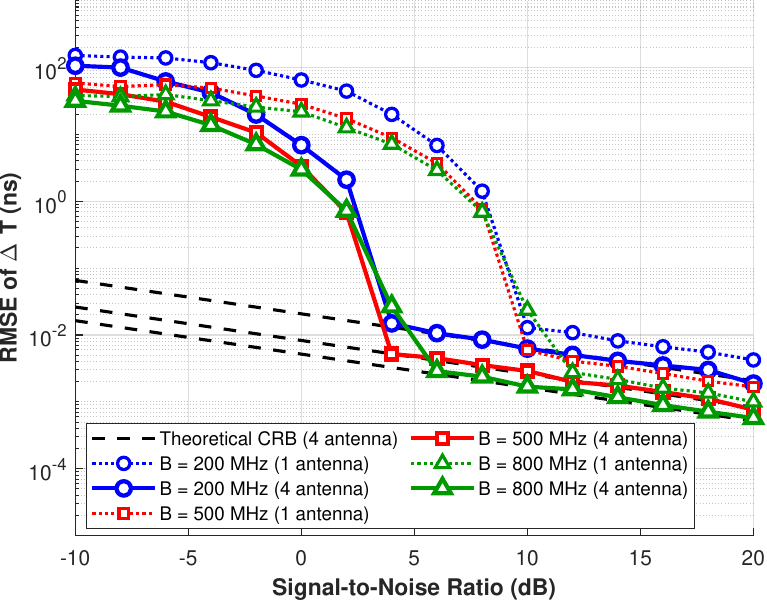}\label{fig:fig10a}}
    \hfill
    \subfloat[Phase calibration RMSE.]{\includegraphics[width=0.48\linewidth]{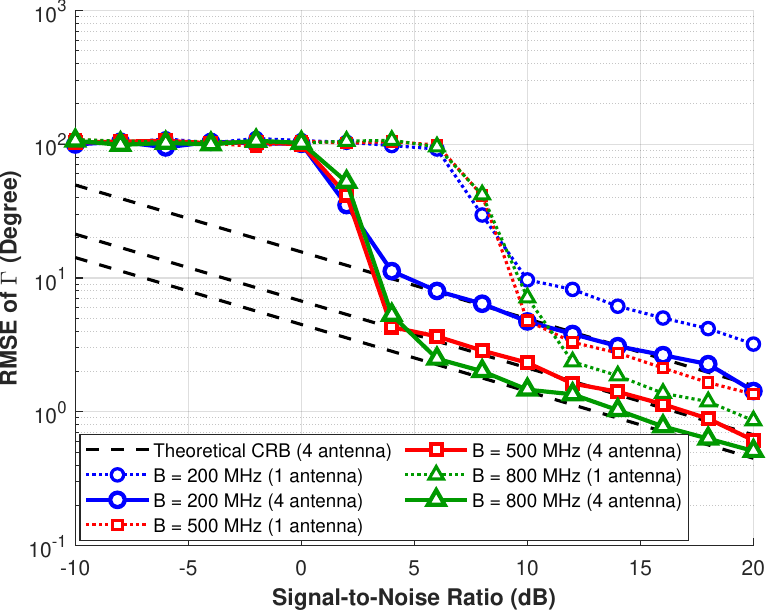}\label{fig:fig10b}}
    \caption{The impact of bandwidth on performance.}
    \label{fig:fig10}
\end{figure}

As shown in Fig. \ref{fig:fig10}, the simulation results present the RMSE of the parameter estimation alongside their corresponding theoretical bounds under varying bandwidth conditions. As the signal bandwidth expands from $200\text{ MHz}$ to $800\text{ MHz}$, the variance of the instantaneous frequency increases, which monotonically lowers the theoretical bounds for both the clock and phase parameters, indicating that a larger bandwidth intrinsically provides stronger theoretical estimation capabilities. The proposed algorithm utilizing the 4-antenna strictly coincides with the theoretical bounds for both parameters, maintaining this optimal accuracy down to an SNR convergence threshold of approximately $4\text{ dB}$. Conversely, the single-antenna exhibits degraded performance, with its convergence threshold delayed to $10\text{ dB}$. Furthermore, even in the converged high-SNR regime, the single-antenna scheme exhibits a consistent performance gap, yielding errors approximately half an order of magnitude higher than the multi-antenna-based bounds due to the absence of local spatial coherent integration gain.

\subsection{Experiment 6: Comprehensive Performance Comparison with Traditional Baselines}
\label{subsec:exp6}

This experiment comprehensively compares the proposed GHR framework with traditional baseline algorithms to evaluate its relative superiority in realistic calibration scenarios. For the distributed receiver network, a baseband equivalent model is utilized with a sampling rate of $5\text{ GHz}$. The network consists of a reference node and an uncalibrated node, each equipped with a 4-element sub-array. The macroscopic distance between the distributed nodes is set to approximately $3$ meters. For the far-field transmitting node, the cooperative calibration source emits an LFM signal. The carrier frequency is $2\text{ GHz}$, the signal pulse duration is $1\ \mu\text{s}$, and the bandwidth is $500\text{ MHz}$. The SNR is swept from $-10\text{ dB}$ to $20\text{ dB}$, with 200 independent Monte Carlo trials performed at each level to obtain statistically stable RMSEs for both clock and phase estimations. The baseline methods selected for comparison include a physical-layer generalized cross-correlation method with sub-sample interpolation and a MAC-layer ordinary least squares protocol based on two-way message exchanges. Traditional "two-step" phase calibration is applied to these baselines using their respective time delay estimates.

\begin{figure}[t]
    \centering
    \subfloat[Time calibration RMSE.]{\includegraphics[width=0.48\linewidth]{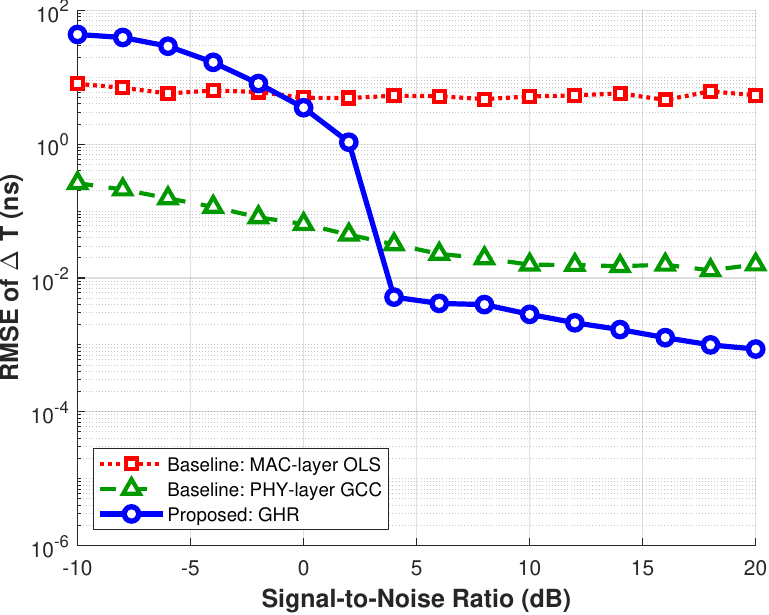}\label{fig:fig11a}}
    \hfill
    \subfloat[Phase calibration RMSE.]{\includegraphics[width=0.48\linewidth]{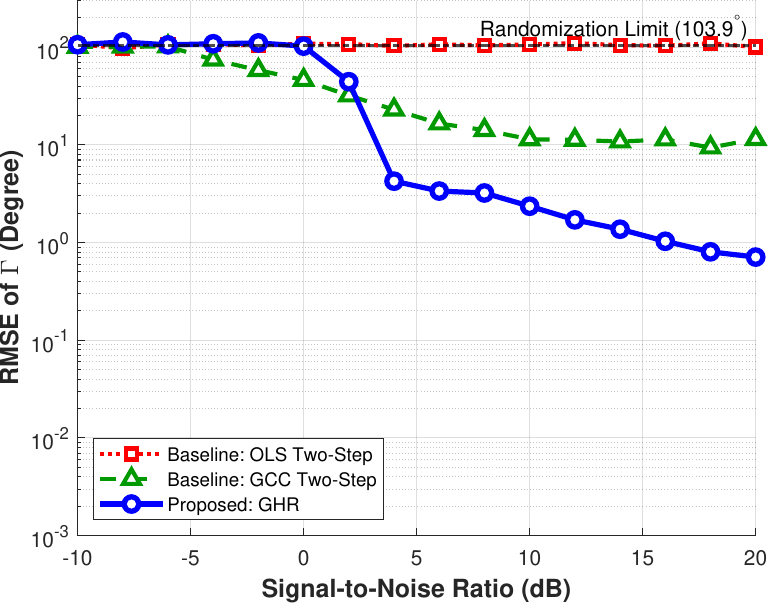}\label{fig:fig11b}}
    \caption{Comparative simulation.}
    \label{fig:fig11}
\end{figure}

As presented in Fig. \ref{fig:fig11}, the simulation results show the RMSE for the clock delay and initial phase estimations. Regarding the clock delay estimation, the proposed scheme converges rapidly above a $4\text{ dB}$ SNR threshold, reaching an accuracy level of approximately $10^{-3}\text{ ns}$. The cross-correlation baseline exhibits a steady and uniform decline across the spectrum. While it performs slightly better than the proposed algorithm in the extremely low-SNR regime, its performance becomes an order of magnitude worse than the proposed scheme at high SNRs due to the discrete interpolation error floor. Considering that distributed calibration is inherently a cooperative scenario typically operating in high-SNR environments, this performance gap clearly highlights the superiority of the proposed framework. The ordinary least squares baseline performs the worst, maintaining a stable error at the nanosecond level regardless of the signal quality due to inherent MAC-layer queuing uncertainties. 

Regarding the phase estimation, the proposed algorithm again converges rapidly above $4\text{ dB}$, yielding an accuracy an order of magnitude higher than the cross-correlation method. The ordinary least squares baseline completely fails to resolve the phase, stabilizing at the theoretical random limit of $103.9^\circ$ due to severe phase cycle wrapping caused by its nanosecond-level time residuals. This demonstrates the fundamental collapse of the traditional two-step calibration paradigm and validates the necessity of the proposed joint orthogonal decoupling.

\section{Conclusion}

This paper resolves the severe time-phase coupling and spatial aliasing challenges for multi-node cooperative sensing in large-aperture distributed IoT networks. By exploiting the high-order kinematic evolution of non-stationary signals, we propose the GHR framework. This mathematical framework deterministically flattens the highly twisted spatial phase trajectory onto a multidimensional dynamic hyper-plane, establishing a multivariate regression model that orthogonally decouples macroscopic clock offsets and microscopic RF phase errors. To guarantee practical deployment on low-cost, resource-constrained IoT edge nodes, we establish a passive feature-level distributed calibration architecture. Operating via unidirectional feature-level transmission, this architecture completely eliminates MAC-layer uncertainties and bidirectional packet exchanges. Furthermore, by adopting a LFM source to achieve dimensionality reduction, the framework plummets the algorithmic complexity to a strict linear order, ensuring ultimate engineering feasibility. Extensive evaluations demonstrate that the proposed framework achieves robust, closed-form picosecond-level synchronization under severely degraded SNR conditions across kilometer-scale IoT apertures.


\appendix[Derivation of the Cram\'er-Rao Bound]
\label{app:crb}

\setcounter{equation}{0}
\renewcommand{\theequation}{A.\arabic{equation}}

In this appendix, we derive the theoretical Cram\'er-Rao Bound (CRB) for the joint estimation of the macroscopic clock synchronization error $\Delta T_m$ and the microscopic RF phase error $\Gamma_m$ within the proposed GHR framework.

According to the dimensional-reduction decoupling mechanism (Section III-B), for an LFM calibration source, the unwrapped phase difference trajectory over $K$ discrete snapshots can be modeled as a linear observation equation:
\begin{equation}
    \Delta\Psi_m(t_k) = -\big(\tau_m(\theta) + \Delta T_m\big) \omega(t_k) + \tilde{\Gamma}_m + e_m(t_k)
    \label{eq:app_obs}
\end{equation}
where $\tilde{\Gamma}_m = \Gamma_m + \pi\mu\big(\tau_m(\theta) + \Delta T_m\big)^2$ is the generalized intercept, and $e_m(t_k)$ represents the residual phase measurement noise. Assuming the noise is zero-mean AWGN with variance $\sigma_{\Psi}^2$. Thanks to the local coherent integration over a sub-array of $M_{sub}$ antennas, the equivalent phase noise variance is substantially suppressed, i.e., $\sigma_{\Psi}^2 \propto 1 / (M_{sub} \cdot \text{SNR})$.

Let the unknown parameter vector for the $m$-th uncalibrated node be defined as $\boldsymbol{\Theta}_m = [\Delta T_m, \tilde{\Gamma}_m]^T$. The log-likelihood function of the observation sequence $\boldsymbol{\psi}_m = [\Delta\Psi_m(t_1), \dots, \Delta\Psi_m(t_K)]^T$ is given by:
\begin{equation}
\begin{aligned}
\ln \mathcal{L}(\boldsymbol{\Theta}_m)
&= - \frac{K}{2}\ln(2\pi\sigma_{\Psi}^2) \\
&\quad - \frac{1}{2\sigma_{\Psi}^2}
\sum_{k=1}^K \Big( \Delta\Psi_m(t_k)
- \hat{\Psi}_m(t_k, \boldsymbol{\Theta}_m) \Big)^2
\end{aligned}
\end{equation}
where $\hat{\Psi}_m(t_k, \boldsymbol{\Theta}_m) = -(\tau_m(\theta) + \Delta T_m)\omega(t_k) + \tilde{\Gamma}_m$.

The elements of the $2 \times 2$ Fisher Information Matrix (FIM), denoted as $\mathbf{I}(\boldsymbol{\Theta}_m)$, are defined as:
\begin{equation}
    [\mathbf{I}(\boldsymbol{\Theta}_m)]_{i,j} = -E\left[ \frac{\partial^2 \ln \mathcal{L}(\boldsymbol{\Theta}_m)}{\partial \Theta_i \partial \Theta_j} \right]
\end{equation}

Computing the second-order partial derivatives yields:
\begin{align}
    \frac{\partial^2 \ln \mathcal{L}}{\partial \Delta T_m^2} &= -\frac{1}{\sigma_{\Psi}^2} \sum_{k=1}^K \omega^2(t_k) \\
    \frac{\partial^2 \ln \mathcal{L}}{\partial \tilde{\Gamma}_m^2} &= -\frac{K}{\sigma_{\Psi}^2} \\
    \frac{\partial^2 \ln \mathcal{L}}{\partial \Delta T_m \partial \tilde{\Gamma}_m} &= \frac{1}{\sigma_{\Psi}^2} \sum_{k=1}^K \omega(t_k)
\end{align}

To simplify the notation, we define the mean instantaneous frequency $\bar{\omega} = \frac{1}{K}\sum_{k=1}^K \omega(t_k)$ and the mean square frequency $\overline{\omega^2} = \frac{1}{K}\sum_{k=1}^K \omega^2(t_k)$. The FIM can then be compactly written as:
\begin{equation}
    \mathbf{I}(\boldsymbol{\Theta}_m) = \frac{K}{\sigma_{\Psi}^2} 
    \begin{bmatrix}
        \overline{\omega^2} & -\bar{\omega} \\
        -\bar{\omega} & 1
    \end{bmatrix}
\end{equation}

The CRB matrix is the inverse of the FIM, $\mathbf{C} = \mathbf{I}^{-1}(\boldsymbol{\Theta}_m)$. The determinant of the FIM is:
\begin{equation}
    |\mathbf{I}| = \frac{K^2}{\sigma_{\Psi}^4} (\overline{\omega^2} - \bar{\omega}^2) = \frac{K^2}{\sigma_{\Psi}^4} \text{Var}(\omega)
\end{equation}
where $\text{Var}(\omega)$ represents the temporal variance of the instantaneous frequency, which intrinsically characterizes the dynamic bandwidth of the non-stationary signal.

Inverting the FIM, we obtain the theoretical lower bounds for the clock synchronization error and the generalized phase error:
\begin{align}
    \text{CRB}(\Delta T_m) &= [\mathbf{C}]_{11} = \frac{\sigma_{\Psi}^2}{K \cdot \text{Var}(\omega)} \label{eq:crb_time} \\
    \text{CRB}(\tilde{\Gamma}_m) &= [\mathbf{C}]_{22} = \frac{\sigma_{\Psi}^2 \cdot \overline{\omega^2}}{K \cdot \text{Var}(\omega)} = \frac{\sigma_{\Psi}^2}{K} \left( 1 + \frac{\bar{\omega}^2}{\text{Var}(\omega)} \right) \label{eq:crb_phase}
\end{align}

These closed-form expressions reveal two profound physical insights regarding dynamic manifold synchronization:
\begin{enumerate}
    \item Equation \eqref{eq:crb_time} proves that the estimation accuracy of the macroscopic clock error is strictly inversely proportional to the dynamic variance of the signal's instantaneous frequency ($\text{Var}(\omega)$). If a stationary monochromatic signal is utilized ($\text{Var}(\omega) = 0$), the CRB approaches infinity, mathematically confirming the Dynamic Observability Condition proposed in Section IV-C.
    \item Since the true RF phase error $\Gamma_m$ is extracted by subtracting a known deterministic quadratic constant from $\tilde{\Gamma}_m$, its theoretical variance fundamentally obeys \eqref{eq:crb_phase}. It demonstrates that the phase estimation accuracy is governed by the ratio of the mean carrier frequency to the dynamic bandwidth ($\bar{\omega}^2 / \text{Var}(\omega)$). This perfectly explains the theoretical performance divergence (error floor tendency) observed in large-scale apertures when the fractional bandwidth is insufficient.
\end{enumerate}

\bibliographystyle{IEEEtran}
\bibliography{refs}

\end{document}